\documentclass[twocolumn,floatfix, showpacs]{revtex4}

\usepackage[final]{epsfig}
\usepackage{amsmath}
\usepackage{parskip}
 
\begin{document}
 
\title{Note on proton-antiproton suppression in 200 AGeV Au-Au collisions}
 
\author{Thorsten Renk}
\email{trenk@phys.jyu.fi}
\author{Kari J.~Eskola}
\email{kari.eskola@phys.jyu.fi}
\affiliation{Department of Physics, P.O. Box 35 FI-40014 University of Jyv\"askyl\"a, Finland}
\affiliation{Helsinki Institute of Physics, P.O. Box 64 FI-00014, University of Helsinki, Finland}
 
\pacs{25.75.-q,25.75.Gz}
\preprint{HIP-2006-46/TH}

\begin{abstract}
We discuss the measured nuclear suppression of $p+ \overline{p}$ production in 200 AGeV Au-Au collisions at RHIC within radiative energy loss. For the AKK set of fragmentation functions, proton production is dominated by gluons, giving rise to the expectation that the nuclear suppression for $p+ \overline{p}$ should be stronger than for pions due to the stronger coupling of gluons to the quenching medium. Using a hydrodynamical description for the soft matter evolution, we show that this is indeed seen in the calculation. However, the expected suppression factors for pions and protons are sufficiently similar that a discrimination with present data is not possible. In the high $p_T$ region above 6 GeV where the contributions of hydrodynamics and recombination to hadron production are negligible, the model calculation is in good agreement with the data on $p+ \overline{p}$ suppression.
\end{abstract}
 
\maketitle

The STAR collaboration has measured the nuclear suppression ratio $R_{CP}$ (yield for $0-12$\% central collisions divied by yield for $60-80$\% peripheral collisions) for $\pi^+ + \pi^-$ and $p + \overline{p}$ in 200 AGeV Au-Au collisions \cite{STAR-data}, finding that the suppression pattern of both pions and protons is very similar in the $6+$ GeV momentum region. This is to some degree surprising, as the AKK fragmentation functions \cite{AKK} (which roughly describe the STAR data on proton production in p-p collisions \cite{STAR-pp-pp}) indicate that $\sim 80$ \% of $p+ \overline{p}$ production comes from gluon jets whereas only $\sim 40$ \% of pion production is gluon-driven. This is not so in the older KKP set of fragmentation functions \cite{KKP} which consequently underpredict proton production at STAR. However, gluons as colour octet states are expected to undergo stronger energy loss in medium than quarks, thus the dominance of gluon jets in hard proton production should map into a stronger suppression of $p + \overline{p}$ production as compared to $\pi^+ + \pi^-$ production. 

It becomes thus a quantitative question in the energy loss model to check how large this difference in suppression should be and if present data are able to resolve it.  A framework to study energy loss within full expansion dynamics of soft matter has been presented in \cite{Correlations} and extended to a 3-d hydrodynamical model \cite{Hydro} in \cite{Jet3d}. We utilize this framework to address the above question. We stress that given the scenario describing pionic $R_{AA}$, the change to study the suppression of protons is only the trivial change of the fragmentation function, no further parameters are adjusted. Furthermore, since the AKK fragmentation function overpredicts $p+ \overline{p}$ production in the high $p_T$ region in p-p collisions (and the fragmentation function of $q(\overline{q}) \rightarrow p(\overline p)$ is well constrained), thus overemphasizing the role of gluons, we stress that our results are an upper limit for the expected difference between proton and pion suppression.

Let us now discuss the treatment of partons propagating through the medium: 
our calculation follows the BDMPS formalism for radiative energy loss 
\cite{Baier:1996sk} using quenching weights as introduced by
Salgado and Wiedemann \cite{Salgado:2002cd,Salgado:2003gb}.
The probability density $P(x_0, y_0)$ for finding a hard vertex at the 
transverse position ${\bf r_0} = (x_0,y_0)$ and impact 
parameter ${\bf b}$ is given by the product of the nuclear profile functions as
\begin{equation}
\label{E-Profile}
P(x_0,y_0) = \frac{T_{A}({\bf r_0 + b/2}) T_A(\bf r_0 - b/2)}{T_{AA}({\bf b})},
\end{equation}
where the thickness function is given in terms of Woods-Saxon the nuclear density
$\rho_{A}({\bf r},z)$ as $T_{A}({\bf r})=\int dz \rho_{A}({\bf r},z)$. 
If we call the angle between outgoing parton and the reaction plane $\phi$, 
the path of a given parton through the medium $\xi(\tau)$ is specified 
by $({\bf r_0}, \phi)$ and we can compute the energy loss 
probability $P(\Delta E)_{path}$ for this path. We do this by 
evaluating the line integrals
\begin{equation}
\label{E-omega}
\omega_c({\bf r_0}, \phi) = \int_0^\infty \negthickspace d \xi \xi \hat{q}(\xi) \quad  \text{and} \quad \langle\hat{q}L\rangle ({\bf r_0}, \phi) = \int_0^\infty \negthickspace d \xi \hat{q}(\xi)
\end{equation}
along the path where we assume the relation
\begin{equation}
\label{E-qhat}
\hat{q}(\xi) = K \cdot 2 \cdot \epsilon^{3/4}(\xi) (\cosh \rho - \sinh \rho \cos\alpha)
\end{equation}
between the local transport coefficient $\hat{q}(\xi)$ (specifying 
the quenching power of the medium), the energy density $\epsilon$ and the local flow rapidity $\rho$ with angle $\alpha$ between flow and parton trajectory \cite{Flow1,Flow2}.
Here $\omega_c$ is the characteristic gluon frequency, setting the scale of the energy loss probability distribution, and $\langle \hat{q} L\rangle$ is a measure of the path-length weighted by the local quenching power.
We view 
the parameter $K$ as a tool to account for the uncertainty in the selection of $\alpha_s$ and possible non-perturbative effects increasing the quenching power of the medium (see discussion in \cite{Correlations}) and adjust it such that pionic $R_{AA}$ for central Au-Au collisions is described.

Using the numerical results of \cite{Salgado:2003gb}, we obtain $P(\Delta E; \omega_c, R)_{path}$ 
for $\omega_c$ and $R=2\omega_c^2/\langle\hat{q}L\rangle$ as a function of jet production vertex and the angle $\phi$.

From the energy loss distribution given a single path, we can define the averaged energy loss probability distribution $P(\Delta E)\rangle_{T_{AA}}$ as
\begin{equation}
\label{E-P_TAA}
\langle P(\Delta E)\rangle_{T_{AA}} \negthickspace = \negthickspace \frac{1}{2\pi} \int_0^{2\pi}  
\negthickspace \negthickspace \negthickspace d\phi 
\int_{-\infty}^{\infty} \negthickspace \negthickspace \negthickspace \negthickspace dx_0 
\int_{-\infty}^{\infty} \negthickspace \negthickspace \negthickspace \negthickspace dy_0 P(x_0,y_0)  
P(\Delta E)_{path}.
\end{equation}
The energy loss probability  $P(\Delta E)_{path}$ is derived in the limit of infinite parton energy  \cite{Salgado:2003gb}. In order to account for the finite energy of the partons we truncate $\langle P(\Delta E) \rangle_{T_{AA}}$ 
at $\Delta E = E_{\rm jet}$ and add $\delta(\Delta-E_{\rm jet}) \int^\infty_{E_{\rm jet}} d\epsilon P(\epsilon)$ to the truncated distribution to ensure proper normalization. The physics meaning of this correction is that we consider all partons as absorbed whose energy loss is formally larger than their initial energy.
We calculate the momentum spectrum of hard partons in leading order perturbative QCD (LO pQCD) (explicit expressions are given in \cite{Correlations} and references therein). The medium-modified perturbative production of hadrons can then be computed from the expression
\begin{equation}
d\sigma_{med}^{AA\rightarrow h+X} \negthickspace \negthickspace = \sum_f d\sigma_{vac}^{AA \rightarrow f +X} \otimes \langle P(\Delta E)\rangle_{T_{AA}} \otimes
D_{f \rightarrow h}^{vac}(z, \mu_F^2)
\end{equation} 
with $D_{f \rightarrow h}^{vac}(z, \mu_F^2)$ the fragmentation function with momentum fraction $z$ at scale $\mu_F^2$ \cite{AKK}, and from this we compute the nuclear modification factor $R_{AA}$  as
\begin{equation}
R_{AA}(p_T,y) = \frac{dN^h_{AA}/dp_Tdy }{T_{AA}({\bf b}) d\sigma^{pp}/dp_Tdy}.
\end{equation}

\begin{figure}
\epsfig{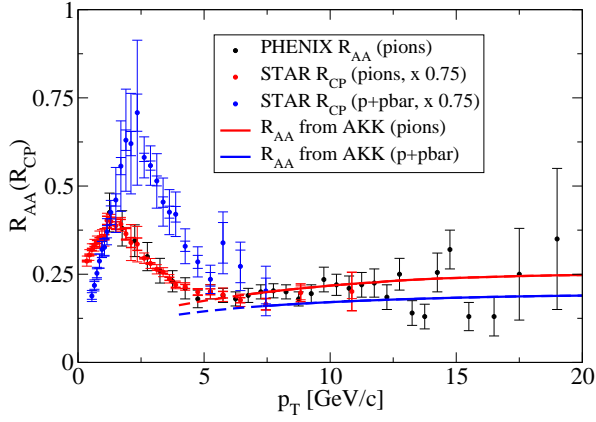}
\caption{\label{F-1}$R_{AA}$ for pions as measured by PHENIX \cite{PHENIX_R_AA}, scaled $R_{CP}$ for pions measured by STAR \cite{STAR-data} (in good agreement with the PHENIX $R_{AA}$), scaled $R_{CP}$ for protons measured by STAR \cite{STAR-data} and calculated $R_{AA}$ for pions and protons, dashed lines indicate the regime where recombination is expected to become an important effect \cite{Reco}.}
\end{figure} 

In Fig.~\ref{F-1} we compare with the STAR data. Since our model, making use of thermodynamical quantities to describe energy loss induced by the soft medium is not suited to describe $60-80$\% peripheral collisions where a dilute, predominantly hadronic, medium is expected to exist, we calculate $R_{AA}$ and scale the experimental data. We show that the scaled pionic $R_{CP}$ as obtained by STAR \cite{STAR-data} is in perfect agreement with the pionic $R_{AA}$ as obtained by PHENIX \cite{PHENIX_R_AA}. This gives us confidence that the comparison of protonic $R_{AA}$ with scaled $R_{CP}$ is not grossly unreasonable.
As apparent from the results, the similarity of the pionic and protonic suppression is expected, the difference between pion and proton $R_{AA}$ is less than the error on the pion $R_{AA}$ over a large kinematic range.

Let us briefly discuss the origin of this outcome: While for an infinite energy parton the average energy loss scales linearly with the coupling strength to the medium (and hence with the color charge), $R_{AA}$ for finite $p_T$ cannot exhibit a linear scaling downward for the simple reason that it is bounded from below by zero. In other words, once the medium quenching is strong enough such that essentially all partons from a spacetime region are absorbed, increasing the quenching power of this region further cannot alter $R_{AA}$ any more and consequently observed hadrons in such a situation are emitted from the medium surface (cf. the discussion in \cite{Dainese}). In the present calculation, we likewise observe this saturation effect (albeit somewhat weaker, cf. \cite{Correlations}), predominantly for gluons.

\begin{figure}
\epsfig{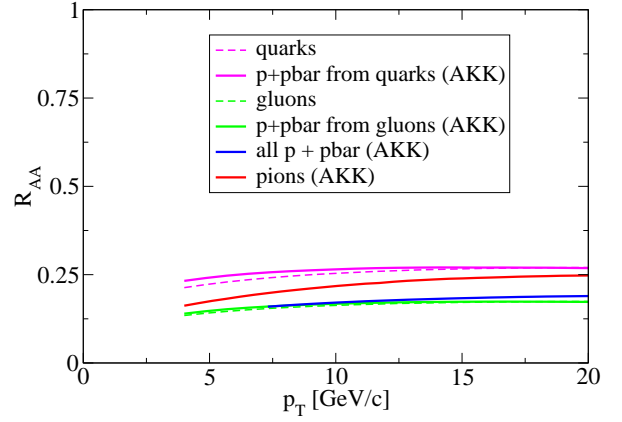}
\caption{\label{F-2}Calculated $R_{AA}$ for quarks and gluons and protons/antiprotons originating from quarks and gluons using the AKK fragmentation functions as well as the final result averaged over the relative contributions of quarks and gluons in the pQCD calculation.}
\end{figure} 

In Fig.~\ref{F-2} we show $R_{AA}$ for quarks and gluons as well as $R_{AA}$ for protons produced exclusively from quarks and gluons. Since the fraction of protons produced from quarks is small when using the AKK fragmentation functions, the final result for protons is very close to the gluon $R_{AA}$ at all accessible $p_T$ (whereas the pion $R_{AA}$ is between the two partonic curves, showing the transition from a gluon-dominated parton spectrum at low $p_T$ to a quark-dominated spectrum at higher $p_T$). The gluonic $R_{AA}$ is clearly in a saturated regime and we checked that the spread between quark and gluon $R_{AA}$ is reduced by a factor two as compared to a medium with less quenching power in which $R_{AA}$ for pions is close to 0.5. It is also apparent that while the functional form of the fragmentation function has some influence on the results at lower $p_T$, $R_{AA}$ can mainly be understood from the partonic suppression and the $p_T$ dependent ratio of quarks and gluons contributing to the production of a given hadron. Thus, the similarity of $R_{AA}$ for protons and pions in this calculation has to be attributed to the fact that gluon emergy loss appears in the saturated regime.

Once again, we stress that we consider the calculation to be an upper limit for the expected difference. For the most part, the STAR data on proton suppression are in a regime where hydrodynamics or recombination  \cite{Reco} is the dominant contribution to hadron production, in the momentum region above 6 GeV where fragmentation is expected to dominate hadronization, the calculation is in agreement with the data.

\begin{acknowledgments}

We thank Jan Rak and Hannu Paukkunen for valuable comments and discussions. This work was financially supported by the Academy of Finland, 
Project 115262. 
 
\end{acknowledgments}

\end{document}